\documentclass[runningheads]{llncs}
\usepackage{cite}
\usepackage{amsmath,amssymb,amsfonts}
\usepackage{algorithm}
\usepackage{algorithmicx}
\usepackage[noend]{algpseudocode}
\usepackage[caption=false]{subfig}

\usepackage{graphicx}
\usepackage{textcomp}
\usepackage{xcolor}
\usepackage{tcolorbox}
\usepackage{graphicx}

\usepackage{hyperref}

\usepackage{comment}
\usepackage{todonotes}

\usepackage{latexsym}
\usepackage{subfig}
\usepackage{float}

\usepackage{colortbl}
\usepackage{multirow}
\usepackage{adjustbox}

\usepackage{chngcntr}
\usepackage{booktabs}
\usepackage{enumitem}
\usepackage{xurl}

\graphicspath{{figs/}}

\begin{document}
\title{Business Process Simulation with Differentiated Resources: Does it Make a Difference?}
\titlerunning{Business Process Simulation with Differentiated Resources}
%
\author{Orlenys L\'opez-Pintado \and
Marlon Dumas}
\authorrunning{O. L\'opez-Pintado and M. Dumas}
%
\institute{University of Tartu, Tartu, Estonia \\
\email{\{orlenyslp, marlon.dumas\}@ut.ee}}
\maketitle              
\begin{abstract}
Business process simulation is a versatile technique to predict the impact of one or more changes on the performance of a process.
Mainstream approaches in this space suffer from various limitations, some stemming from the fact that they treat resources as undifferentiated entities grouped into resource pools. 
These approaches assume that all resources in a pool have the same performance and share the same availability calendars.
Previous studies have acknowledged these assumptions, without quantifying their impact on simulation model accuracy.
This paper addresses this gap in the context of simulation models automatically discovered from event logs.
The paper proposes a simulation approach and a method for discovering simulation models, wherein each resource is treated as an individual entity, with its own performance and availability calendar. An evaluation shows that simulation models with differentiated resources more closely replicate the distributions of cycle times and the work rhythm in a process than models with undifferentiated resources. 
\end{abstract}

\section{Introduction}
\label{sec:introduction}

Business Process (BP) simulation~\cite{Aalst15} is a technique to analyze ``what-if'' scenarios, such as ``what would be the cycle time of a process if the number of daily new cases increases by 20\%?'' (\textbf{S1}) or ``what if two resources involved in a process become unavailable for an extended period of time?'' (\textbf{S2}). 

The starting point for BP simulation is a simulation model consisting of a process model enhanced with parameters capturing the available resource capacity, activity processing times, arrival rate of new cases, etc.
It has been noted that existing BP simulation approaches suffer from various limitations~\cite{Nakatumba10,Aalst15,Freitas15}. 
Some of these limitations stem from incompleteness of, or inaccuracies in, the BP simulation model.
These limitations are partly addressed by data-driven simulation methods~\cite{MartinDC16,CamargoDG20}, which automatically discover and calibrate simulation models from execution data (\emph{event logs}). These methods ensure that the simulation model is better aligned with the observed reality~\cite{Aalst15,MartinDC16,CamargoDG20}. 
Other limitations of BP simulation approaches relate to assumptions made by the underlying BP simulator~\cite{Aalst15,Freitas15}, most notably the assumption that resources are interchangeable entities. Specifically, mainstream BP simulation approaches, including data-driven ones, make the following assumptions:
\begin{itemize}
    \item[\textbf{A1}] \emph{Pooled resource allocation}. Each resource belongs to one resource pool (e.g., a role or group). Resource pools are disjoint. All instances of an activity are allocated to the same resource pool. For example, all instances of tasks \emph{Check invoice} and \emph{Schedule payment} are allocated to an \emph{Accountant} pool.
    \item[\textbf{A2}] \emph{Undifferentiated performance}. The processing time of an activity does not depend on the resource who performs it.
    \item[\textbf{A3}] \emph{Undifferentiated availability}. All resources in a pool are available for work during the same time periods, e.g., Monday to Friday, 9:00-17:00.
\end{itemize}

In practice, each (human) resource has their own capabilities, performance, and availability. Previous studies have hypothesized that the above assumptions affect the accuracy of simulation models~\cite{Nakatumba10,Aalst15,Freitas15,AfifiAA18}, but without quantifying their impact.
In this setting, this paper addresses the following question: \emph{Do assumptions \textbf{A1-A3} affect the accuracy of a business process simulation model, and if so, to what extent?}
The paper studies this question in the context of simulation models discovered from event logs.
To address this question, the paper proposes and evaluates: (1) a business process simulation approach with differentiated resources; and (2) an automated method to discover a simulation model with differentiated resources from an event log.
In the proposed approach, resources are not grouped into pools, but  treated as individuals (unpooled allocation), 
the performance of each resource is independent of that of other resources (differentiated performance), and each resource may have its own availability calendar (differentiated availability).
As a result, a simulation model can be used not only to answer what-if scenarios \textbf{S1} and \textbf{S2} above, but also scenarios such as: ``what if resource R is replaced by resource R$'$ with lower performance?'' (\textbf{S3}) or ``what if a resource changes their availability from full-time to part-time?'' (\textbf{S4}).


The paper is structured as follows. Sect.~2 discusses related work. Sect.~3 formalizes assumptions \textbf{A1}-\textbf{A3} by presenting a simulation approach with undifferentiated resources. Sect.~4 presents a simulation approach with differentiated resources, while Sect~5 proposes a corresponding method to discover simulation models. Sect.~6 empirically compares simulation models with differentiated vs.\ undifferentiated resources, and Sect.~7 concludes and sketches future work.

\section{Related Work}
\label{sec:related}

Van der Aalst et al.~\cite{Nakatumba10,Aalst15} analyze three limitations of BP simulation approaches: unreliability of simulation models for short-term prediction, insufficient reliance on execution data to construct simulation models, and incorrect modeling of resources. The authors emphasize that resources often work part-time and that failure to capture this, leads to inaccurate simulations. In~\cite{NakatumbaWA12}, the authors study the impact of workload on resource performance, i.e., to what extent resource performance varies depending on workload and the impact of this variability on simulation accuracy. Our contribution is related to these studies, but we focus on limitations that arise when resources are modeled as undifferentiated entities.

Afifi et al.~\cite{AfifiAA18} note that existing BP simulation approaches, including the BPSim simulation modeling standard~\cite{bpsim}, rely on role-based resource allocation, and do not support a wider range of resource allocation styles such as those identified in~\cite{RussellAHE05}. However, the authors do not quantify the impact of the identified limitations (e.g., role-based allocation) on concrete simulation scenarios.



Freitas \& Pereira~\cite{Freitas15} reviews five BP simulation tools. They find that these tools do not allow one to define unavailability periods for individual resources. However, they do not evaluate the impact of this limitation. Some commercial simulation engines such as IBM Websphere Modeler\footnote{\url{https://www.ibm.com/support/pages/download-websphere-business-modeler-advanced-v70}} support the definition of ``named resources'', which can have their own timetables (differentiated availability). However, the activity processing times are defined at the level of tasks, and hence they do not support differentiated performance.



This paper studies the impact of resource differentiation on simulation models discovered from logs. Prior studies on BP simulation model discovery~\cite{RozinatMSA09,MartinDC16,CamargoDG20} assume that resources are available 24/7. In~\cite{Estrada-TorresC21}, the authors address this limitation by integrating a technique for discovering timetables into a simulation model discovery pipeline, assuming all resources in a pool have the same timetable.




\section{Simulation Models with Undifferentiated Resources}
\label{sec:non-diff-simulation}

A BP simulation model with pooled allocation and undifferentiated resources (herein, a \emph{classic BP simulation model}) consists of a process model $M$ (e.g., a BPMN diagram) enhanced with simulation metadata  described in Def.~\ref{def:non-diff}. 



\begin{definition}[Classic BP Simulation Model]\label{def:non-diff}
A classic BP simulation model is a tuple $<E, A, G, F, RPools, Alloc, PT, BP, AT, AC>$, where $E, A, G$ are respectively the sets of events, activities, and gateways of a BPMN model, $F$ is the set of directed flow arcs of a BPMN model, and the remaining elements capture simulation parameters as follows:
 \begin{enumerate}
     \item $RPools$ is a set of resource pools. Each resource pool $p \in RP$ represents a group of resources. The resource pools are disjoint, i.e., $\forall\ p_1, p_2 \in RPools\ : p_1 \cap p_2 = \emptyset$. Each resource pool is described by the following properties:
     \begin{itemize}
         \item {\sc Size(p)} $\in \mathbb{N}$ is the number of resources in the pool.
         \item {\sc Avail(p)} is a calendar (a set of intervals) during which every resource in $p$ is available to perform activity instances.
         \item {\sc Cost(p)} is the cost of each pool $p$ per time unit (e.g., hour).
     \end{itemize}
     \item {\sc Alloc} $: A \rightarrow RP$ is a function mapping each activity $a \in A$ to one resource pool $p \in RPools$. A resource pool can perform many activities.
     \item $PT : A \rightarrow \mathcal{P}(\mathbb{R}+)$ is a mapping from each activity $a \in A$ to a probability density function, modeling the the processing times of activity $a$.
     \item {\sc BP}$ : F \rightarrow [0, 1]$ is a function that maps each flow $f \in F$ s.t., the source of $f$ is an element of $G$ to a probability (a.k.a., the branching probability).
     \item {\sc AT} $\in \mathcal{P}(\mathbb{R}+)$ is a probability density function modeling the inter-arrival times between consecutive case creations. 
     \item {\sc AC} is calendar (set of intervals) such that cases can only be created during an interval in {\sc AC}.
 \end{enumerate}
\end{definition}



Given that in classic BP simulation models, resource pools are disjoint, they cannot capture scenarios where participants share their time across multiple pools (cf.\ assumption \textbf{A1} in Sect.~1). Also, since all resources in a pool have the same timetable, these models cannot capture scenarios where a pool incorporates some part-time resources and some full-time ones (assumption \textbf{A3}). Finally, in classic BP simulation models, the processing times of an activity do not depend on the resource that performs it. Hence, such models cannot capture scenarios where some resources in a pool are faster or slower than others (assumption \textbf{A2}).

When executed in a simulation engine, a (classic) BP simulation model produces an event log as per  Def.~\ref{def:event-log}. Herein, we call \emph{simulated logs} those logs produced by a simulation and \emph{real logs} those extracted from information systems.

\begin{definition}[Event log]\label{def:event-log}
 An event log $E$ is a set of events, each representing the execution of an activity instance in a process. An event $e \in E$ is a tuple $e = <\alpha, r, \tau_0, \tau_s, \tau_c>$, where $\alpha$ is the label of one activity in a business process (i.e., $e$ is an instance of the activity $\alpha$), $r$ is the resource who performed $\alpha$, $\tau_0$ is the timestamp in which the activity instance was enabled to be executed, and $\tau_s$, $\tau_c$ are, respectively, the timestamps corresponding to the beginning and end of the activity instance. A trace (a.k.a., process case) is a non-empty sequence of events $t = <e_1, e_2, ..., e_n>$, and an event log $L = <t_1, t_2, ..., t_m>$ is a non-empty sequence of traces, each capturing one instance of a process (i.e., a case). 
\end{definition}

 Various performance metrics can be computed from a log, including: \textit{waiting time} -- the time-span from the moment the activity is enabled until the starting of the corresponding event; \textit{processing time} -- the time-span between beginning and end of the event; \textit{cycle time} -- the difference between the end time and start time of a case; and \textit{resource utilization} -- the ratio between the time a resource is busy executing activity instances, divided and its total availability time. 

\section{Simulation Models with Differentiated Resources}
\label{sec:approach}

To lift the limitations imposed by assumptions \textbf{A1-A3} (cf.\ Sect. 1), we propose an approach to BP simulation with differentiated resources. In this simulation model, the notion of \emph{resource pool} is replaced by that of \emph{resource profile}. Like a resource pool, a resource profile models a set of resources that share the same availability calendar. However, unlike classic BP simulation models, an activity in a process model may be assigned to multiple resource profiles and the same resource profile may be shared by multiple pools. For example, in a claims handling process, there may be a resource profile for \emph{junior claims handler}, another for \emph{senior claims handler} and a third for \emph{lead claims handler}, each with different calendars. Activity \emph{Analyze claim} may be assigned to \emph{junior claims handler} and \emph{senior claims handler}, i.e.,  an instance of \emph{Analyze claim} may be performed by a junior or by a senior claims handler. Meanwhile, activity \emph{Assess claim} may be assigned to \emph{senior claims handler} and \emph{lead claims handler}. Finally, activity \emph{Approve large claim} may be assigned to \emph{lead claims handler}, i.e., only lead claims handlers may perform this activity. 
Another difference is that in a classic simulation model, each activity is mapped to a distribution of processing times. Meanwhile, in a simulation model with differentiated resources, the distribution of processing times depends not only on the activity, but also on the resource profile. Thus, the distribution of processing times of the activity {\it Analyze claim} when assigned to a {\it junior claims handler} is different than when assigned to a {\it senior claims handler}, e.g., seniors may be faster, on average, than juniors.



\begin{definition}[BP simulation model with differentiated resources]\label{def:diff}
A BP simulation model with differentiated resources $DSM$ is a tuple $<E, A, G, F,$ {\sc RProf}$, BP, AT, AC>$, where $E, A, G$ are the sets of events, activities, and gateways of a BPMN model, $F$ is the set of directed flow arcs of a BPMN model, and the remaining elements capture simulation parameters as follows:   
 \begin{enumerate}
     \item {\sc RProf} $= \{r_1, ..., r_n \}$ is a set of resource profiles, where $n$ is the number of resources in the process, and each resource $r \in R$ is described by:
     \begin{itemize}
         \item {\sc Alloc} $(r) =  \{\alpha\ |\ \alpha \in A\}$ is the set of activities that $r$ can execute, 
         \item {\sc Perf} $(r, \alpha)= R \times A^m \rightarrow \mathcal{P}^{m}(\mathbb{R}+)$ is a mapping from the resource $r$ to a list of density functions over positive real numbers, corresponding to the distribution of processing times of each activity $\alpha \in$ {\sc Alloc}, with $m$ being the number of activities that $r$ can perform,
         \item {\sc Avail}($r$) is the calendar (a set of intervals) in which the resource $r$ is available to perform each activity $\alpha \in$ {\sc Alloc}, 
         \item {\sc Cost}($r$) is the cost of the resource $r$ per time unit (e.g., hour)
     \end{itemize}
     \item {\sc BP}, {\sc AT}, and {\sc AC} are defined as in Def.~\ref{def:non-diff}.
 \end{enumerate}
\end{definition}


The key difference between Def.~\ref{def:diff} and Def.~\ref{def:non-diff} is that instead of mapping each activity to a pool, Def.~\ref{def:non-diff} maps each resource profile to the set of activities, and for each activity, it  captures the corresponding probability density function of processing times. Note that a classic simulation model can be converted into a model with differentiated resources by mapping each resource pool to one resource profile. However, a scenario where  an  activity is assigned to multiple resource profiles cannot be captured as a classic simulation model. Note also that if every resource profile has a size of one (i.e., one profile per resource), each resource may have different performance and availability. In Sect.~\ref{sec:discovery}, we focus on discovering such models with individualized resources.

The operational semantics of simulation models with differentiated resources is captured by 
Alg.~\ref{algo:approach}. This algorithm takes as input a simulation model $DSM$ according to Def.~\ref{def:diff}, the number $pCases$ of process instances to simulate, and the timestamp $startAt$ of the beginning of the simulation. Like in a classic BP simulation engine, the simulation produces a log and the performance indicators in Sect.~\ref{sec:non-diff-simulation}. Due to space limitations, we illustrate steps related to the generation and update of the simulation events, focusing on the functions in Def.~\ref{def:diff}, but omitting the details of the data structures and algorithms required to handle the event logs, calendars, scheduling, and estimation of performance indicators. 
\begin{algorithm}[tp]
\begin{algorithmic}[1]
\scriptsize
\Function{SimulateProcess}{$DSM$, $pCases$, $startAt$}
\For{{\bf each} resource $r$ $\in$ $DSM$}
\State $readyAt[r]$ $\gets$ $minFrom(${\sc Avail} $, startAt)$
\EndFor
\State {\tt diffResQ} $\gets$ {\sc DiffResourceQueue} ({\sc Alloc}, {\sc Avail}, {\sc SortingCriteria}= $min(readyAt)$)
\State {\tt evtQ} $\gets$ {\sc GenerateAllArrivalEvents} ($pCases$, $DSM$, {\sc AT}, {\sc AC})
\While{{\tt evtQ} {\bf not empty}}
\State $e$ $\gets$ {\sc PopEvent}({\tt evtQ})
\State $e[r]$ $\gets$ {\sc PopResource}({\tt diffResQ}, $e[\alpha]$)
\State $e[\tau_s]$ $\gets$ $\max(e[\tau_0], readyAt[e[r]])$
\State $e[\tau_c]$ $\gets$ $e[\tau_s]$  + {\sc IdleProcessingTime} ($e[\tau_s]$, $e[r]$, $e[\alpha]$, {\sc Avail}, {\sc Perf})
\State $readyAt[e[r]]$ $\gets$ $e[\tau_c]$ + $IdleTime(r,$ {\sc Avail}, $e[\tau_c])$
\State {\sc UpdateResourceAvailability}({\tt diffResQ}, $e[r]$)
\State {\sc UpdateSimulatedEventLog}(e)
\State $state$, $enabled$ $\gets$ {\sc UpdateProcessState}($e[\alpha]$, $e[pState]$, $DSM$, {\sc BP})
\For {{\bf each} $\alpha'$ $\in$ $enabled$}
\State $nE$ $\gets$ $Event(\alpha = \alpha', \tau_0 = e[\tau_c], pState = state)$
\State {\sc EnqueueEnabledEvent}({\tt evtQ}, $nE$)
\EndFor
\EndWhile
\EndFunction
\end{algorithmic}
\caption{Snippet of processes simulation with differentiated resources}
\label{algo:approach}
\end{algorithm}

 The first issue to handle in models with differentiated resources is that they can be shared among several tasks. Unlike undifferentiated models, which allow only one pool per activity, multiple resource profiles may be allocated to each activity in differentiated scenarios. To address this, we use a multi-queue data structure named {\sc DiffResourceQueue}, initialized in line 4. The queue groups the resources by activities according to function {\sc Alloc}, restricting allocated resources to the remaining shared activities. Besides, resources are sorted in the queue according to a priority function {\sc SortingCriteria} given as input. By default, the resource sorting criteria consider the minimum timestamp in which each resource will be ready to perform an activity, i.e., stored in the map $readyAt$. Thus, the values in the map $readyAt$ (initialized in lines 2-3) are calculated considering the resources working calendars, given by the function {\sc Avail}, and the periods in which resources are busy performing activities during the simulation. The support for multiple sorting criteria in {\sc DiffResourceQueue} opens many options for prioritizing and sorting resources following different criteria, e.g., allocate resources according to their expertise given some conditions. 
 
 Next, function {\sc GenerateAllArrivalEvents} in line 5 produces the initial event (see Def.~\ref{def:event-log}) of each process case to simulate, i.e., according to the arrival time distribution {\sc AT}, in the intervals defined by the arrival calendar {\sc AC}. The queue {\tt evtQ} stores and retrieves all the simulated events according to the timestamp in which the corresponding activity $\alpha$ was enabled. Then the simulation proceeds until there is not enabled event in {\tt evtQ} (line 6). We are using the notation $e[r]$, $e[\alpha]$, $e[\tau_0]$, $e[\tau_s]$ and  $e[\tau_c]$ referring respectively to the resource allocated, activity name, enabling, starting and completing times of the event $e$ (see Def.~\ref{def:event-log}). Additionally, $e[pState]$ represents the marking over the flow-arcs of the corresponding process instance at the moment of the event creation. This marking simulates the token game as specified in the BPMN standard. For each process instance created by the function {\sc GenerateAllArrivalEvents}, it generates tokens that traverse the flow-arcs in the model until reaching the end event in the BPMN model. An element in the control flow becomes enabled when one or many tokens arrive at its incoming flow-arcs (i.e., according to the element execution semantics). Similarly, the execution of an enabled element consumes the incoming tokens, generating new ones on its outgoing flow-arcs.  


The queue {\tt evtQ} only stores enabled events. Thus, the attributes $e[r]$, $e[\tau_s]$ and $e[\tau_c]$ are determined and updated once the corresponding event is popped from {\tt evtQ}, i.e., the event is then executed. In lines 7-8 of Alg.~\ref{algo:approach}, the event with the lowest enabling timestamp in {\tt evQ} is allocated to a resource, according to availability and allocation criteria passed to the resources queue {\tt diffResQ}, i.e., selecting the participant being available the earliest as default criteria. 

When the event is enabled, the allocated resource may not be according to their calendar (and vice-versa). Thus, the starting timestamp of the event relies on both task and resource availability (line 9). Next, in line 10, the completion timestamp is calculated by the function {\sc IdleProcessingTime} which adjusts the ideal processing time (if the resource works in the task without interruption according to {\sc Perf}), plus the time the resource may rest from their calendar in {\sc Avail}. Similarly, function {\sc IdleTime} calculates the next timestamp the resource is available after completing the task, updating the resource queue accordingly (lines 11-12). Finally, lines 14-17 update the process state, retrieving the activities enabled after executing the current event, queuing them as events in {\tt evtQ} with enabling time equal to the completion time of the previous event.
  
\section{Discovering Differentiated Resources Profiles}
\label{sec:discovery}

This section proposes an approach to discover simulation models with differentiated resources described in Sec.~\ref{sec:approach}. Due to space limitations, we focus only on the main steps to discover differentiated resource profiles from event logs, i.e., to model each resource performance and availability independently. Before describing our proposal, Def.~\ref{def:calendar} formalizes the weekly calendars, followed by Def.~\ref{def:notation} introducing some notations we will use across this section.   


\begin{definition}\label{def:calendar}
A weekly calendar $\widehat{C}$ is binary relation $W \times \Delta$ between the set of weekdays, $W = \{Monday, ..., Sunday\}$, and a set of time granules $\Delta = \{\delta_1, ..., \delta_n\}$ where $\bigcap_{i=1}^{n}\delta_{i} = \emptyset$. Each time granule $\delta_i \in \Delta$ is a sorted pair of time points $<\tau^w_s, \tau^w_c>$, such that $\tau^w_s, \tau^w_c = <hour, minute, second>$, $hour \in [0, ..., 23]$, $minute, second \in [0, ..., 60]$, and $\tau^w_s \leq \tau^w_c$. A calendar entry $\kappa$ is a tuple $<\omega, \tau^w_s, \tau^w_c>$ representing a time interval for a given day. For example, $\kappa = <$Monday, 08:15:00, 12:00:00$>$ describes Monday from 08:15 to 10:30.
\end{definition}

\begin{definition}[Notations]\label{def:notation}
\begin{itemize}
    
    
    
    
    \item Given an event log $L$: $E$ is the set of all the events in $L$, $R$ and $A$ are, respectively, the sets of resources and activities in any event $e \in E$. Besides, $A_r = \{ \alpha \in A\ |\ \exists\ e \in E, r \in R, \alpha \in A : r, \alpha \in e \}$, and {\bf $E_r$}$ = \{ e \in E\ |\ r \in R \wedge r \in e \}$ are the set of activities and events executed by the resource $r$, respectively. With, $E_\alpha = \{ e \in E\ |\ \alpha \in A \wedge \alpha \in e \}$ being the set of events, which are instances of the activity $\alpha$, and $E_{r,\alpha} = \{ e\ |\ e \in E_r \cap E_\alpha \}$ the set of instances of $\alpha$ executed by the resource $r$.
    
    \item $\Gamma$ is function mapping a timestamp in the event log into a calendar entry $\kappa = <\omega, \tau^w_s, \tau^w_c>$, where $<\tau^w_s, \tau^w_c>$ spans $n$ minutes. Specifically, $\Gamma$ retrieves an interval of size $n$ containing the timestamp received as input.  Note that, $\Gamma$ retrieves intervals assuming that days are split into intervals of equal size $n$ starting from the $00:00:00$ hours, e.g., from n=15 min days are split as $[00:00:00 - 00:15:00), [00:15:00 - 00:30:00), ..., [23:45:00 - 00:00:00)$. For example, consider a calendar with time intervals of 15 minutes, for the timestamp $2022-01-01 T 08:12$, the function $\Gamma$ returns the calendar entry candidate $<Saturday, <08,00,00>, <08,15,00>>$.

    
    
    
    
    \item $\Omega^n_r = \{\kappa^{m(\kappa)}\ |\ \forall <\tau_s, \tau_c>\ \in E_r,\ n > 0,\ \kappa = \Gamma(\tau_s, n) \wedge \kappa = \Gamma(\tau_c, n)\}$ is a multi-set of calendar entry candidates of duration $n$ mapped from the starting and ending timestamps of each event executed by the resource $r$, with the supra-index $m(\kappa)$ being the number of calendar entries $\kappa$ in $\Omega^n_r$.
    
    \item $\Omega^n_{r, \alpha} = \{\kappa^{m}\ |\ \kappa \in \Omega^n_r \wedge \alpha \sim \kappa,\ n > 0\}$ is the subset of $\Omega^n_r$ containing all the calendar entry candidates that are instances of the activity $\alpha$, with $\sim$ representing that an instance of $\alpha$ occurred in the calendar entry $\kappa$.
    
    
\end{itemize}
\end{definition}

To discover resource availability calendars, we take inspiration from the approach in~\cite{LiWJ00}, which discovers repetition patterns from a set of time granules with a certain level of confidence and support. The latter approach assumes time intervals that are covered entirely. This condition does not hold when discovering working intervals of a resource, since the event log shows only the start and completion timestamps of each event, and gives no information about what happens in two timestamps. Also, the start of an event is conditioned by the enablement of the related activity, i.e., a resource can be available but still needs to wait to start an activity until it becomes enabled in the process. Thus, we redefined the confidence and support metrics in~\cite{LiWJ00} to discover calendars over time granules not fully described by the input data. Furthermore, we filter the resources with low frequency according to their relative participation, to exclude external resources (i.e., resource who seldom participate in the process), as there is insufficient data to discover availability calendars for such resources individually.
 
Other approaches such as~\cite{MartinDCS20} can be used to discover resource availability calendars. In this latter work, the authors use the activity waiting and processing times of each activity to estimate the intervals resources are available according to an input event log. Thus,~\cite{MartinDCS20} assumes that available resources will work as soon as an activity is enabled, and they keep working during the entire activity's execution interval (without any break). In this paper, do not make this assumption. Instead, we only assume that the resource was working when the activity instance starts and when it completes, and in-between, we consider that the resource may or may not be working. In any case, the calendar discovery approach in~\cite{MartinDCS20} could be used as an alternative to the approach presented here.

 
 
 Defs.~\ref{def:conf},~\ref{def:supp}, and~\ref{def:part} describe, respectively, the metrics of confidence, support, and resource participation we use to filter and discover the resource profiles. The metrics retrieve a real number between 0 (worst assessment) and 1, the best possible value. The activity-conditional confidence, given a calendar entry $\kappa = <\omega, \tau^w_s, \tau^w_c>$ related to an activity $\alpha$, measures the ratio between the number of times $\alpha$ was started or completed on the weekday $\omega$ between $\tau^w_s$ and $\tau^w_c$, divided by the total of weekdays $\omega$ that $\alpha$ occurred. For example, it measures from every Monday a resource was observed executing a given activity, how often it happened between 8:00 AM - 8:15 AM. Def.~\ref{def:conf} generalizes the metric to a set of tasks executed by a resource in the same time granules as the maximum between the individual value computed for each activity. The support metric computes from all the timestamps a resource was active in the log, what ratio is covered by some calendar entry. Finally, the participation metric estimates the ratio of events performed by a resource compared with the number of events executed by the most frequent resource. The comparison is relative to the activities each resource can perform. For example, resources $r_1$ and $r_2$ may execute 10 and 1000 events, respectively. If we compare $r_1$ and $r_2$ globally, then $r_1$ has a participation ratio of 0.01 compared to $r_2$. However, if $r_1$ and $r_2$ execute different activities, and if $r_1$ is the only executing all the instances of an activity, the relative ratio is 1.0 as $r_1$ is relevant to the activity $r_1$ performs alone. 

\begin{definition}\label{def:conf}
$ {\tt Confidence}(r, \kappa) = \frac{\underset{\alpha \in A_r, \alpha \sim \kappa}{\max} |\Omega^n_{r, \alpha}|}{|\{\omega^m\ |\ \omega \in W\ \wedge\ \omega \in \Omega^n_{r, \alpha} \} |}$ computes the activity-conditioned confidence of a calendar entry $\kappa = <\omega, \tau^w_s, \tau^w_e>$. The multi-set in the fraction denominator computes how many times each activity $\alpha$ was executed on the weekday $\omega$.
\end{definition}

\begin{definition}\label{def:supp}
 ${\tt Support}(r, \widehat{C}) = \frac{|\{\kappa^m\ |\ \kappa \in \Omega^n_r \wedge \kappa \in \widehat{C} \}|}{|\Omega^n_r|}$ computes the support of a given calendar $\widehat{C}$, where the multi-set in the fraction numerator computes how many calendar entries $\kappa$  from the multi-set of candidates $\Omega^n_r$ are covered by $\widehat{C}$.
\end{definition}

\begin{definition}\label{def:part}
${\tt RParticipation(r)} = \frac{\sum_{\alpha \in A_r}|E_{r,\alpha}|}{\sum_{\alpha \in A_r}\underset{r' \in R}{\max} | \{E_{r',\alpha}\}|}$ computes the relative participation of a respurce $r$. The fraction numerator computes the number of events executed by $r$. The denominator sums up all the events executed by each resource who executed the most events for each activity executed by $r$.
\end{definition}

\begin{algorithm}[tp]
\begin{algorithmic}[1]
\scriptsize
\Function{DiscoverResourceProfiles}{$L$, $DSM$, $n$, $dSupp$, $dConf$, $dPart$}
\State {\sc ParseEventLog($L$)}  \Comment{To extract sets and multi-sets in Def.~\ref{def:notation}} 
\State {\sc Alloc}, {\sc Avail}, {\sc Perf}  $\gets$ $\emptyset$, $\emptyset$, $\emptyset$
\For {{\bf each} $r$ $\in$ $R$}
    \State $\Omega^n_r$ $\gets$ {\sc ExtractCalendarEntries}($E_r$, $\Gamma$, $n$)
    \If{{\sc RParticipation}($r$) $\geq$ $dPart$ }
        \State {\sc Avail}$[r]$ $\gets$ {\sc DiscoverCalendar}($\Omega^n_r$, $dSupp$, $dConf$)
    \Else
        \State {\sc Avail}$[r]$ $\gets$ $\emptyset$
    \EndIf
\EndFor

\For {{\bf each} $\alpha$ $\in$ $A$}
    \State $discarded$ $\gets$ $\emptyset$
    \For {{\bf each} $r$ $\in$ $R$ : {\sc Avail}$[r]$ $=$ $\emptyset$ {\bf and} $E_{r,\alpha}$ $\neq$ $\emptyset$}
        \State $discarded$.{\sc Add}($E_{r,\alpha}$)
    \EndFor
    \State $jointR$ $\gets$ {\sc MaxDisjointIntervals($discarded$)}
    \For{{\bf each} $r$ $\in$ $jointR$}
        \State $\Omega^n_r$ $\gets$ {\sc ExtractCalendarEntries}($jointR$, $\Gamma$, $n$)
        \State $\widehat{C}$ $\gets$ {\sc DiscoverCalendar}($\Omega^n_r$, $dSupp$, $dConf$)
        \If{$\widehat{C}$ $\neq$ $\emptyset$}
            \State {\sc Avail}$[r]$.{\sc Add}($\widehat{C}$)
        \EndIf
    \EndFor
    \If{{\sc IsUnallocated}($\alpha$)}
        \State {\sc BuildUnrestrictedCalendar}($jointR$, {\sc Avail})
    \EndIf
\EndFor
\For{{\bf each} $r$ $\in$ {\sc Avail} : {\sc Avail}[$r$] $\neq$ $\emptyset$}
    \State {\sc Alloc}$[r]$ $\gets$ $A_r$
\EndFor
\For{{\bf each} $\alpha$ $\in$ $A$}
    \State  {\sc Perf}[$\alpha$].{\sc Add}({\sc DiscoverProcessingTimes}($E_\alpha$, $R$))
\EndFor
\State \Return {\sc Alloc}, {\sc Avail}, {\sc Perf}
\EndFunction

\end{algorithmic}
\caption{Resource Profiles Discovery (from event logs)}
\label{algo:discovery}
\end{algorithm}

\begin{algorithm}[tp]
\begin{algorithmic}[1]
\scriptsize
\Function{DiscoverCalendar}{$\Omega^n_r$, $dSupp$, $dConf$}
\State $\widehat{C}$, $discarded$ $\gets$ $\emptyset$, $\emptyset$
\For {{\bf each} $<\omega, \tau^w_s, \tau^w_c >$ $\in$ $\Omega^n_r$}
    \If{{\tt Confidence}($<\omega, \tau^w_s, \tau^w_c>$, $\Omega^n_r$) $\geq$ $dConf$}
        \State $\widehat{C}$.{\sc Add}($<\omega, \tau^w_s, \tau^w_c>$)
    \Else
        \State $discarded$.{\sc Add}($<\omega, \widehat{\tau_s}, \widehat{\tau_c} >$)
    \EndIf
\EndFor

\If{{\tt Support}($\widehat{C}$, $\Omega^n_r$) $<$ $dSupp$} 
    \State {\sc SortMultisetByMultiplicity}($discarded[r]$, order={\tt decreasing})
    \For{$<\omega, \tau^w_s, \tau^w_c >$ $\in$ $discarded$}
        \State $\widehat{C}$.{\sc Add}($<\omega, \tau^w_s, \tau^w_c>$)
        \State $discarded$.{\sc Remove}($<\omega, \tau^w_s, \tau^w_c>$)
        \If{{\sc Support}($\widehat{C}$, $\Omega^n_r$) $\geq$ $dSupp$}
            \State {\bf break}
        \EndIf
    \EndFor
\EndIf
\State \Return $\widehat{C}$
\EndFunction

\end{algorithmic}
\caption{Calendar Discovery}
\label{algo:calendar}
\end{algorithm}

\begin{algorithm}[tp]
\begin{algorithmic}[1]
\scriptsize
\Function{DiscoverProcessingTimes}{$E_\alpha$, $R$, $binSize=50$}
\State $\widehat{D}$ $\gets$ $\emptyset$

\State $pendingResources$ $\gets$ $\emptyset$
\For {{\bf each} $r$ $\in$ $R$}
    \If{$|E_{r,\alpha}|$ $\geq$ $binSize$}
        \State $\widehat{D}[r]$ = {\sc BestFittedDistribution}($E_{r,\alpha}$, {\sc Alloc}, $binSize$)
    \Else 
        \State $pendingResources$.{\sc Add}($r$)
    \EndIf
\EndFor

\State $jointD$ $\gets$  {\sc BestFittedDistribution}($E_\alpha$, $binSize$)

\For {{\bf each} $r$ $\in$ $pendingResources$}
    \State $\widehat{D}[r]$ $\gets$ $jointD$
\EndFor
\State \Return $\widehat{D}$

\EndFunction

\end{algorithmic}
\caption{Processing Time Distribution Discovery}
\label{algo:distribution}
\end{algorithm}

 Alg.~\ref{algo:discovery} captures the main steps to calculate differentiated resource profiles. It takes as input an event log, a BPMN model, the size $n$ of the granules in the calendar, the desired support, confidence, and participation values, and the minimum number of data points required to infer the processing-time distributions. Line 2 extracts from the log the sets and multi-sets described in Def.~\ref{def:notation}, followed by the initialization of the mappings {\sc Alloc}, {\sc Avail} and {\sc Perf} in Def.~\ref{def:diff}. Lines 4-9 discard the resources with low relative participation (Def.~\ref{def:part}), storing (in the mapping {\sc Avail}) the discovered calendars of each resource over the required threshold. Function {\sc ExtractCalendarEntries}, in line 5 transforms the timestamps in which each resource was active into calendar entries according to $\Gamma$ (cf.\  Def.~\ref{def:notation}). Function {\sc DiscoverCalendar} in line 7 is described by Alg.~\ref{algo:calendar}.

 To discover a calendar, Alg.~\ref{algo:calendar} receives a multi-set of calendar entry candidates of a given resource $r$. Then, lines 3-7 iterate over each candidate, adding those with confidence above $dConf$ in the calendar $\widehat{C}$, discarding the remaining ones. Next, line 8 verifies if the calendar achieved the required support $dSupp$. If not, the algorithm adds the most frequent entries until reaching the required support (lines 9-14). Thus, the algorithm relies on confidence only to filter potential outliers among the entry candidates, prioritizing that the calendar always covers the ratio of timestamps described by the support. 

 Filtering the resource and calendar entries in lines 6-7 of Alg.~\ref{algo:discovery} may cause the coverage of some tasks to become too low. As a result, an activity that is executed rarely or that is executed by external resources (i.e., resources from outside the organization, who seldom participate in the process) can lose all their resources, if none of them fulfills the participation threshold. This issue is addressed by Alg.~\ref{algo:discovery} in lines 10-21 by grouping the events of the removed resources related to each activity and assigning them to \emph{aggregated resources}.
 Function {\sc MaxDisjointIntervals} takes those grouped events and: (1) Sort them in ascending order of their start times $\tau_s$, (2) add event $e'$ with the highest $\tau_s$, deleting all events whose time interval intersects $e'$, (3) repeat (1)-(2) until no intervals remain. Next, an aggregated resource is created from each set of events retrieved. The calendar of the aggregated resource is built from the maximal set of mutually disjoint time intervals~\cite{AgarwalKS98}, i.e., by grouping the calendar entries that were discarded due to low confidence. 
 Then, lines 15-19 create a calendar for each aggregated resource. If none fulfills the confidence and support requirement, lines 20-21 retrieve a single calendar as an aggregation of all the discarded events of the related activity without checking for confidence and support values.
 
 Lines 22-23 of Alg.~\ref{algo:discovery} allocate, to each discovered resource, the activities executed by them in the event log. Then, function {\sc DiscoverProcessingTimes} (line 25) estimates the differentiated resource performance as described in Alg.~\ref{algo:distribution}, which from every pair activity resource (lines 4-5), validates the number of events extracted fulfills a certain level of significance $binSize$ (above 50 by default). Resources below the threshold $binSize$ are grouped, with their performance discovered as an aggregation of all their events (lines 7-11). Function {\sc BestFittedDistribution} adjusts each event duration by the calendar of the corresponding resource. Then, it builds a histogram from the event durations and applies curve-fitting to find a probability distribution, from a library of distributions, that best approximates the histogram (the one with lowest residual sum).

\section{Implementation and Evaluation}
\label{sec:evaluation}

 We implemented the proposed approach as an open-source (Python-based) simulation engine, namely {\sc Prosimos}, available at \url{https://github.com/AutomatedProcessImprovement/Prosimos}. {\sc Prosimos}  supports the simulation of processes with an unpooled allocation model and differentiated availability and performance as per Sect.~\ref{sec:approach}. Besides, it provides a component to automatically discover a simulation model with differentiated resources from an event log, as described in Sect.~\ref{sec:discovery}. {\sc Prosimos} takes as input a BPMN process model with simulation parameters as per Def.~\ref{def:diff} (encoded in JSON format). Like other simulation engines, {\sc Prosimos} produces an event log and a set of performance indicators such as waiting, processing, and cycle times, and resource utilization. 

 Using {\sc Prosimos}, we conducted an empirical evaluation aimed at answering the following sub-questions derived from the question posed in Sect.~\ref{sec:introduction}: {\bf EQ1} What impact does unpooled resource allocation have compared to pooled allocation? {\bf EQ2} What impact does differentiated resource performance have compared to undifferentiated performance? {\bf EQ3} What impact does differentiated resource availability have compared to undifferentiated availability?

 \vspace{-3.5mm}
\paragraph{Datasets}
 We use five simulated (synthetic) logs and five real-life ones. Since our proposal does not deal with process model discovery, we use the BPMN models generated from the input logs using the Apromore open-source platform,\footnote{\url{https://apromore.com}}, which we manually adjusted to obtain 90\% replay-based fitness. Table~\ref{tbl:log-description} gives descriptive statistics of the employed logs, including number of traces and events and number of activities and resources. Row ``simulation time'' shows the average execution times (in seconds) across five simulation runs.

 \begin{table}[tp]
    \centering
    \small
\begin{adjustbox}{width=0.99\textwidth}
\begin{tabular}{{lcccccccccc}}
    \toprule
     & \textbf{LO-SL/} & \textbf{LO-SH/} & \textbf{LO-ML/} & \textbf{LO-MH/} & \textbf{P-EX/} & \textbf{PRD/}    & \textbf{C-DM/} & \textbf{INS/} & \textbf{BPI-12/}   &  \textbf{BPI-17}\\
    \midrule
    \textbf{Traces}           & 1000  & 1000 & 1000 & 1000 & 608  & 225  & 954  & 1182    & 8616    & 30\,276 \\
    \midrule
    \textbf{Events}           & 9844  & 9782 & 9768	& 9569 & 9119 & 4503 & 4962 & 23\,141 & 59\,302 & 240\,854 \\
    \midrule
    \textbf{Activities}       & 15    & 15   & 15   & 15   & 23   & 23   & 18   & 11      & 8     & 9 \\
    \midrule
    \textbf{Resources}        & 19    & 19   & 34   & 34   & 47   & 54   & 337  & 125     & 68    & 141 \\
    \midrule
    \textbf{Simulation Time}  & 1.27  & 1.24 & 1.25 & 1.24 & 1.07 & 0.72 & 0.73 & 1.29    & 10.32 & 41.97 \\
    \bottomrule
    \end{tabular}
\end{adjustbox}
   \vspace{1.0mm}
   \caption{Characteristics of the business processes used in the experimentation.}
   \label{tbl:log-description}
   \vspace{-9.0mm}
 \end{table}
 
 The first four event logs were obtained by simulating a Loan Origination (LO) process model using Apromore. The model contains 15 tasks assigned to 5 resource pools. We first simulated the model by assigning the same calendar to all resource pools. Using this single-calendar (S) model, we generated two logs: one where the resource utilization of each pool is around 50\% (Low Utilization -- L) and another with a resource utilization of 80\% (High Utilization -- H). The simulation parameters of the H model were identical to the ones of the L model, except that we adjusted the case arrival rate to obtain higher resource utilization. 
 To test the techniques in the presence of multiple calendars, we simulated the same model after assigning different (overlapping) calendars to each of the five resource pools. We simulated this multi-calendar (M) model twice: once with a low utilization (L) and once with high utilization (H). This procedure led to four simulated logs: LO-SL, LO-SH, LO-ML, LO-MH. The fifth log (\textit{purchasing-example (P-EX)}) is part of the academic material of the Fluxicon Disco tool.\footnote{\url{https://fluxicon.com/academic/material/}}

 The first real-life log (\textit{PRD}) is a log of a manufacturing process.\footnote{\url{https://doi.org/10.4121/uuid:68726926-5ac5-4fab-b873-ee76ea412399}}. 
 The second and third are anonymized real-life logs from private processes. The \textit{C-DM} comes from an academic recognition process executed at a Colombian University. The \textit{INS} log belongs to an insurance claims process. The fourth real-life log is a subset of the BPIC-2012 log\footnote{\url{https://doi.org/10.4121/uuid:3926db30-f712-4394-aebc-75976070e91f}} 
 -- of a loan application process from a Dutch financial institution. We focused on the subset of this log consisting of activities that have both start and end timestamps. Similarly, we used the equivalent subset of the BPIC-2017 log\footnote{\url{https://doi.org/10.4121/uuid:5f3067df-f10b-45da-b98b-86ae4c7a310b}}
 , which is an updated version of the BPI-2012 log (extracted in 2017 instead of 2012). We extracted the subsets of the BPI-2012 and BPI-2017 logs by following the recommendations provided by the winning teams of the BPIC-2017 challenge.\footnote{\url{https://www.win.tue.nl/bpi/doku.php?id=2017:challenge}}
 \vspace{-3mm}

\paragraph{Experiment setup and goodness measures}

To address questions {\bf EQ1-EQ3}, we discovered five simulation models from each log using the following approaches:
 \begin{itemize}
     \item {\bf SP-NP-NA} corresponds to an unpooled allocation with undifferentiated performances and availability. We allocate the resources into a single pool, where each resource can execute the same activities as in the log. The resources share an aggregated calendar built from the entire log. The processing time of each activity is discovered by aggregating all its instances without considering the resource who executes them.
     \item {\bf MP-NP-NA} represents a pooled resource allocation with undifferentiated resource profiles. 
     Resources are grouped into disjoint pools assigned to one or several activities according to~\cite{CamargoDG20}. Each resource pool shares a single calendar and shares processing time distribution functions for each related activity, i.e., built by aggregating the events of the resources in the pool.
     \item {\bf MP-DP-NA} is a pooled resource allocation with differentiated performance and undifferentiated availability. We retain the pools and calendars discovered for {\bf MP-NP-NA}. However, we extract differentiated processing time distributions for each pair activity-resource.
     \item {\bf MP-NP-DA} is pooled resource allocation with undifferentiated performance and differentiated availability. We retain the pools and processing-time distributions discovered for {\bf MP-NP-NA}. However, we extract a differentiated calendar from the activity instances of each resource in the pool.
     \item {\bf SP-DP-DA} corresponds to the unpooled resource allocation with differentiated resources and performances proposed in this paper.
 \end{itemize}
 
We assessed the goodness of the discovered models by simulating them using {\sc Prosimos} and measuring the distance between the simulated logs and the original ones. Camargo et al.~\cite{CamargoDG20} propose several measures to assess the goodness of simulation models discovered from data. These measures cover two dimensions: the control-flow and the temporal dimension. 
The techniques proposed in this paper do not affect the control flow. They only deal with resource performance and availability.
Accordingly, we evaluate them using temporal measures. 
In line with~\cite{CamargoDR22}, we compare simulated and real logs by extracting temporal histograms from each log and computing the Earth Movers' Distance (EMD) between these histograms.
We use two EMD metrics, namely {\bf EMD-CT} and {\bf EMD-WR}. 
{\bf EMD-CT} compares the distributions of cycle time of the traces in the logs. This metric captures to what extent the total durations produced by the simulation model resemble those in the real log. To calculate the {\bf EMD-CT}, we group the cycle times in the real log into 100 equidistant bins. Then, we discretize the simulated log by grouping the cycle times of its traces into bins of the same width as those of the real log. We then measure the EMD between these histograms.
The second metric ({\bf EMD-WR}) compares the distribution of timestamps of the events in the two logs. This measure allows us to assess if the simulated and the real log capture similar work rhythms.
To calculate the {\bf EMD-WR}, we transform each log into a histogram by extracting the start and end timestamps of each event in the log, and we group the resulting set of timestamps by hour. We then calculate the EMD between the resulting histograms.

The EMD is defined on an absolute dataset-dependent scale. Thus, EMD distances should not be used to compare the performance of the approach across multiple logs. Below, we use the EMD metrics to assess the relative performance of multiple simulation discovery approaches within a given dataset.
 
The selection of parameters for simulation model discovery may impact the accuracy. Choosing a small granule size, e.g.,  $n = 60$ seconds, may lead to a fragmented calendar with many intervals. Conversely, a large value, e.g., $n = 24$ hours, may lead to unrealistic calendars in which resources are always available. With a low support threshold, the algorithm may discard many timestamps in the log, leading to low coverage of the observed events. To mitigate these issues, we run a grid search over a range of parameters to find a configuration with low confidence (to filter outliers), high support (to cover a representative set of events), and mid-to-low resource participation (to discard resources that rarely participate in the process). The grid search returned a granule size of 60 minutes for all experiments. The confidence values ranged from 0.1 to 0.5, and the support and resource participation ranged between 0.5 and 1.0. 
\vspace{-3mm}

 \paragraph{Results}
 Tables~\ref{tbl:edm-ct} and~\ref{tbl:emd-ed} show the results of the {\bf EMD-CT} and {\bf EMD-WR} metrics, respectively. The results of the {\bf SP-NP-NA} models illustrate that unpooled resource allocations with undifferentiated resource profiles yield, on average, poor results on both metrics. This suggests that undifferentiated availability and performance may lead to less accurate results, especially when resources have considerable differences in availability and performance. Another drawback of this unpooled approach, due to the activities sharing resources, is that resources may become busy executing an activity that they execute rarely. Thus, increasing the waiting times of other shared activities (with higher frequencies) due to the unavailability of the resource. This problem may have more impact on processes with external resources. Still, these unpooled resource allocations with undifferentiated resources may perform well in processes where resources have similar calendars and performance, as shown in the BPI challenge logs.
  
  \begin{table}[tp]
    \centering
    \small
\begin{adjustbox}{width=0.99\textwidth}
\begin{tabular}{{lcccccccccc|c}}
    \toprule
     & \textbf{LO-SL/} & \textbf{LO-SH/} & \textbf{LO-ML/} & \textbf{LO-MH/} & \textbf{P-EX/} & \textbf{PRD/}    & \textbf{C-DM/} & \textbf{INS/} & \textbf{BPI-12/}   &  \textbf{BPI-17} &  \textbf{Mean}\\
    \midrule
    \textbf{SP-NP-NA}  & 4.49 & 3.83 & 15.77 & 35.1  & 17.54 & 21.73 & 10.53 & 11.24 & 10.04 & 3.95 & 13.42 \\
    \midrule
    \textbf{MP-NP-NA}  & 3.77 & 3.58 & 4.64	& 14.44 & 15.11 & 17.2  & 10.53 & 11.28 & 9.99  & 3.94 & 9.45  \\
    \midrule
    \textbf{MP-DP-NA}  & 3.65 & 4.15 & 7.35  & 17.72 & 15.54 & 18.1  & 10.53 & 11.23 & 9.98  & 3.95 & 10.22 \\
    \midrule
    \textbf{MP-NP-DA}  & 4.31 & 6.06 & 4.64  & 8.5   & 10.49 & 18.32 & 10.02 & 11.25 & 6.55  & 3.85 & 8.4   \\
    \midrule
    \textbf{SP-DP-DA}  & \cellcolor{blue!25} 2.19 & \cellcolor{blue!25} 1.82 & \cellcolor{blue!25} 2.44 & \cellcolor{blue!25} 4.9 & \cellcolor{blue!25} 10.26 & \cellcolor{blue!25} 7.32 & \cellcolor{blue!25} 8.83 & \cellcolor{blue!25} 3.33 & \cellcolor{blue!25} 3.84 & \cellcolor{blue!25} 1.32 & \cellcolor{blue!25} 4.63  \\
    \bottomrule
    \end{tabular}
\end{adjustbox}
   \vspace{1.0mm}
   \caption{Results of the {\bf EMD-CT} metric.}
   \label{tbl:edm-ct}
   \vspace{-4.0mm}
 \end{table}
 
   \begin{table}[tp]
    \centering
    \small
\begin{adjustbox}{width=0.99\textwidth}
\begin{tabular}{{lcccccccccc|c}}
    \toprule
     & \textbf{LO-SL/} & \textbf{LO-SH/} & \textbf{LO-ML/} & \textbf{LO-MH/} & \textbf{P-EX/} & \textbf{PRD/}    & \textbf{C-DM/} & \textbf{INS/} & \textbf{BPI-12/}   &  \textbf{BPI-17} &  \textbf{Mean}\\
    \midrule
    \textbf{SP-NP-NA}  & 491.4 & 264.5 & 341.4 & 195.1 & 1728.8 & 511.9 & 302.7 & 9244.1 & \cellcolor{blue!25} 2510.3 & 5177.0 & 2076.7 \\
    \midrule
    \textbf{MP-NP-NA}  & \cellcolor{blue!25} 375.1 & 276.2 & 369.5 & \cellcolor{blue!25} 64.5  & 1755.7 & 518.0 & 254.8 & 9176.4 & 2545.8 & 5141.5 & 2047.8 \\
    \midrule
    \textbf{MP-DP-NA}  & 507.5 & \cellcolor{blue!25} 207.6 & 344.2 & 64.9  & 1722.7 & 447.5 & 266.9 & 9178.5 & 2518.9 & 5134.5 & 2039.3 \\
    \midrule
    \textbf{MP-NP-DA}  & 402.5 & 273.1 & 388.5 & 169.5 & 1807.7 & 467.1 & 347.1 & 9384.5 & 2638.4 & \cellcolor{blue!25} 5129.9 & 2100.8  \\
    \midrule
    \textbf{SP-DP-DA}  & 378.4 & 273.5 & \cellcolor{blue!25} 331.3 & 76.7  & \cellcolor{blue!25} 1692.2 & \cellcolor{blue!25} 216.8 & \cellcolor{blue!25} 238.7 & \cellcolor{blue!25} 8510.9 & 2628.9 & 5277.4 & \cellcolor{blue!25} 1962.5 \\
    \bottomrule
    \end{tabular}
\end{adjustbox}
   \vspace{1.0mm}
   \caption{Results of the {\bf EMD-WR} metric.}
   \label{tbl:emd-ed}
    \vspace{-8.0mm}
 \end{table}
 
 Comparing the pooled models {\bf MP-NP-NA}, {\bf MP-DP-NA}, and {\bf MP-NP-DA} is not straightforward. On average, they exhibited better results than the unpooled and undifferentiated model {\bf SP-NP-NA}. The latter is a consequence of the pooled models preventing the issue of resources allocated to low-frequency tasks (outliers), but at the cost of not modeling processes with resources shared among tasks. Also, in pooled models, the similarity criteria used to group the resources adjust the data points to discover the aggregated calendars and processing time distributions, leading to more accurate approximations. The experiment shows that, on average, the model {\bf MP-NP-DA} gets better values for the {\bf EMD-CT} metric than the models {\bf MP-NP-NA} and {\bf MP-DP-NA}. Suggesting that a pooled model with differentiated availability and undifferentiated performance approximates trace cycle times better than the baseline of pooled allocation with undifferentiated resources. In contrast, the pooled model with undifferentiated availability and differentiated performance {\bf MP-DP-NA} performs better on the metric {\bf EMD-WR} than {\bf MP-NP-DA} and {\bf MP-NP-NA}. 

 As highlighted in Table~\ref{tbl:edm-ct}, the unpooled model {\bf SP-DP-DA} with fully differentiated performance and availability yields the best results w.r.t. metric {\bf EMD-CT}. On average, the values achieved by {\bf SP-DP-DA} are twice better than {\bf MP-NP-NA} and almost three times better than {\bf SP-NP-NA}. This shows that filtering resources with low resource participation (Def.~\ref{def:part}), combined with differentiated modeling of performance and availability, heightens the temporal accuracy of the discovered simulation models. With respect to metric {\bf EMD-WR} (Table~\ref{tbl:emd-ed}), the unpooled models with differentiated resource performance and availability exhibited the best average results. Here, differences are not as significant as with the cycle time estimations. However, unlike with the {\bf EMD-CT}, histograms built for the metric {\bf EDM-WR} are also impacted by the inter-arrival times discovered. For example, assume the discovered inter-arrival intervals would produce more dispersed starting events in the simulation than in the actual process. Consequently, it may lead to a shift in the timestamps of the subsequently simulated events. The {\bf EMD-WR} metric compares the exact timestamps in which each event occurs. Then, a shift of those events may have a more significant impact on the metric evaluation than in the {\bf EMD-CT} metric, which compares the trace durations without taking into account the exact timestamps involved. The inter-arrival time discovery is orthogonal to the primary goal of this paper, thus, kept as future work~\cite{MartinDC15}.~\footnote{We estimate the inter-case arrival distribution by applying curve-fitting to the data series consisting of the start time of each trace. Branching probabilities are estimated by replaying the log over the model and counting the conditional flow traversals.}
 
 To summarize, with respect to question {\bf EQ1}, unpooled models offer the best results. However, as expected, these models perform poorly when the process involves homogeneous resource pools. Regarding questions {\bf EQ2}-{\bf EQ3}, the experiments show that, on average, models with differentiated performance yield better results (w.r.t. replicating the work rhythm) than undifferentiated models. Conversely, models with differentiated availability are able to better replicate the cycle times. If we only take into account one dimension at a time (differentiated performance or availability), we do not observe significant accuracy improvements (w.r.t. to models with undifferentiated resources). Instead, the experiments show that modeling differentiated performance and availability together yield the most visible improvements, both when it comes to replicating the cycle time distribution and the work rhythm.
 \vspace{-3mm}
 
\paragraph{Threats to validity}

The evaluation reported above is potentially affected by the following threats to  validity:  (1) \textbf{\textit{Internal validity}}: the experiments rely only on ten events logs. The results could differ on other datasets. To mitigate this limitation, we selected logs with different sizes and characteristics and from different domains. (2) \textbf{\textit{Construct validity}}: we used two measures of goodness based on histogram abstractions. The results could be different if we employed other measures, e.g.\ similarity measures between time series based on dynamic time warping. (3) \textbf{\textit{Ecological validity}}: the evaluation compares the simulation results against the original log. While this allows us to measure how well the simulation models replicate the as-is process, it does not allow us to assess the accuracy improvements of using differentiated resources in a what-if setting, i.e., predicting the performance of the process after a change.

\vspace{-1mm}
\section{Conclusion}
\label{sec:conclusion}

The paper outlined an approach to discover simulation models where each resource may have its own performance profile (differentiated performance) and its own calendar (differentiated availability). The paper empirically shows that models with differentiated performance and availability produce simulation logs that are closer to the actual logs from which the simulation model is discovered.

The proposal has a few limitations that warrant further research. First, to estimate inter-arrival times, it applies curve-fitting to the data series consisting of the start time of the first activity instance of each trace. However, the actual case creation time may be earlier than the start time of the first activity instance. This limitation may be tackled by using specialized approaches such as the one in~\cite{MartinDC15}.
Second, the approach to discover availability calendars is designed to discover calendars with weekly periodicity. In practice, the availability of a resource may vary across the year (e.g. different availability in summer months than in winter ones), or across a month (e.g., different availability at the start than at the end of a month). Another future work direction is to discover calendars with more complex periodicity.
Third, the approach for calendar  discovery relies on three parameters: confidence, support, and resource participation. In the current implementation, we apply a grid search over narrow parameter ranges to find an optimal configuration. 
Another future work direction is to enhance the approach with a hyperparameter tuning algorithm to explore large configuration spaces.

\smallskip\noindent\textbf{Reproducibility.} The experiments on public datasets may be reproduced by cloning the repository  \url{https://github.com/AutomatedProcessImprovement/Prosimos} (tag \textsf{bpm2022}) and following the instructions given thereon.

\vspace{1.5mm}
\noindent \textbf{Acknowledgment}: Work funded by European Research Council (PIX project).

\bibliographystyle{splncs04}
\bibliography{references}

\end{document}